\documentclass[12pt]{article}

\usepackage{amsmath, amsthm, amssymb, setspace, mathrsfs, color, verbatim, graphicx,hyperref, amsfonts, bm, rotating, lscape, multicol, multirow,bbm,natbib, psfrag, epsf, enumerate} %natbib
\usepackage{soul}
\usepackage[section]{placeins}
\usepackage{float}
\usepackage{caption}
\usepackage{subcaption}
\usepackage{longtable}

\newcommand{\blind}{0}

\newcommand{\widthcoef}{0.68}
\newcommand{\halfwidthcoef}{0.4}

\newtheorem{thm}{Theorem}%[section]

\setcitestyle{aysep={,}}

\begin{document}

\def\spacingset#1{\renewcommand{\baselinestretch}%
{#1}\small\normalsize} \spacingset{1}

%%%%%%%%%%%%%%%%%%%%%%%%%%%%%%%%%%%%%%%%%%%%%%%%%%%%%%%%%%%%%%%%%%%%%%%%%%%%%%

\if0\blind
{
  \title{\bf Band Depth Clustering for Nonstationary Time Series and Wind Speed Behavior}
  \author{Laura L. Tupper\thanks{
    The authors gratefully acknowledge the SOPF working group at Cornell University, led by Ray Zimmerman, Timothy Mount, and Carlos Murillo-Sanchez; and Colin Ponce, Department of Computer Science, Cornell University. Support was provided from Cornell University Institute of Biotechnology and the New York State Division of Science, Technology and Innovation (NYSTAR), a Xerox PARC Faculty Research Award, NSF Grant DMS-1455172, and the Consortium for Electric Reliability Technology Solutions and the Office of Electricity Delivery and Energy Reliability, Transmission Reliability Program of the U.S. Department of Energy.} \hspace{.2cm}\\
    Department of Statistical Science, Cornell University\\
    David S. Matteson\\
    Department of Statistical Science, Cornell University\\
    C. Lindsay Anderson \\
    Department of Biological and Environmental Engineering, Cornell University}
  \maketitle
} \fi

\if1\blind
{
  \bigskip
  \bigskip
  \bigskip
  \begin{center}
    {\LARGE\bf Title}
\end{center}
  \medskip
} \fi

\begin{abstract}

We explore the behavior of wind speed over time, using the Eastern Wind Dataset published by the National Renewable Energy Laboratory. This dataset gives wind speeds over three years at hundreds of potential wind farm sites. Wind speed analysis is necessary to the integration of wind energy into the power grid; short-term variability in wind speed affects decisions about usage of other power sources, so that the shape of the wind speed curve becomes as important as the overall level. To assess differences in intra-day time series, we propose a functional distance measure, the band distance, which extends the band depth of Lopez-Pintado and Romo (2009). This measure emphasizes the shape of time series or functional observations relative to other members of a dataset, and allows clustering of observations without reliance on pointwise Euclidean distance. To emphasize short-term variability, we examine the short-time Fourier transform of the nonstationary speed time series; we can also adjust for seasonal effects, and use these standardizations as input for the band distance. We show that these approaches to characterizing the data go beyond mean-dependent standard clustering methods, such as k-means, to provide more shape-influenced cluster representatives useful for power grid decisions.
\end{abstract}

\noindent%
\vfill
\hfill {\tiny technometrics tex template (do not remove)}

\newpage
%\spacingset{1.45} 

%%%%%%%%%%%%%%%%
%%%% INTRODUCTION %%%%

\section{Introduction}
\label{sec:intro}

A key concern in power engineering is the characterization of the behavior of wind speed (or power output) over time, at a number of wind turbine locations, using only a small number of representatives. These representatives can then used to characterize wind speed behavior for training and testing power system algorithms that incorporate wind generation, such as the SuperOPF (Optimal Power Flow) system described by \cite{murillo-a}, which aims to make optimal decisions about activating generators and dispatching power. The number of representatives is generally constrained by the computational complexity of the algorithm using the data, so we treat this number as fixed in the discussion below.

Wind speed data for thousands of locations are produced by NREL, and available through the EWITS database (see \url{http://www.nrel.gov/electricity/transmission/eastern_wind_dataset.html}). Each site in the database has location information and wind speed measurements at ten-minute time increments for three years, allowing us to consider the data as either a high-frequency time series or functional data. In this analysis, we will look at wind speed in units of days, each day being a single time series from midnight to midnight (whose length is 144, since we use ten-minute time increments). We choose this time frame in the context of making day-ahead decisions about unit commitment (generator activation); but the methods described here can be applied to time series of different length, or different resolution on the time axis, as is desired in some applications. For example, \cite{pourhabib} perform forecasting with a one-hour time resolution and a six-hour window, citing six hours as a typical cutoff for ``short-term" wind forecasts, beyond which meterologically-based models may be preferred to data-driven approaches. We must also consider the nonstationarity of wind speed when viewed as a time series; \cite{pourhabib} take a coarsely-grained approach, defining ``epochs" such as ``6pm to 12 am, all days in January," while we will explore a more high-resolution time-frequency analysis.

\begin{figure}[H]
\centering
\includegraphics[width=\widthcoef\textwidth]{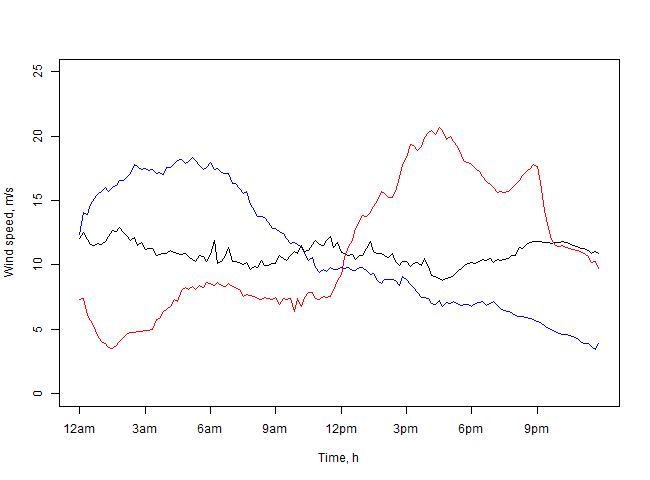}
\caption{Three sample days from the dataset. The mean level for each of these days is approximately 11 m/s, despite their different shapes.}
\label{sam_days}
\end{figure}

An effective set of representative days must cover the range of wind behaviors as far as possible, and include information about the relative probabilities of seeing each type of day. Clustering the data, then, is a natural way to capture different recurrent behaviors: the cluster centers can provide a reasonable representative of each behavior type, while the cluster sizes indicate which types are most likely. But a simple ``high wind, low wind" breakdown, obtained by clustering on average wind level during the day, is largely uninformative. For example, wind power often cannot be stored, and backup generators cannot be activated instantly and have ramping constraints on their output; so the shape of the wind speed curve has a critical effect on optimal decision-making. This shape can vary widely, even among days with similar average levels, as shown in Figure \ref{sam_days}.

Methods for calculating the similarity or distance between two time series have been reviewed in several sources, for example \cite{gunopulos}. Many of these techniques, such as the edit distance or Dynamic Time Warping and its extensions, are designed to allow shifting or stretching behavior along the time axis. If we do not wish to allow time warping, perhaps because we are basing time-sensitive decisions on our results, we may approach the time series as high-dimensional data, with each time point as a dimension. The dimension of a discrete time series is then equal to its length: not as extreme as, for example, some text-matching datasets, but still high enough to require specialized methodology.

\begin{figure}[H]
\centering
\includegraphics[width=\widthcoef \textwidth]{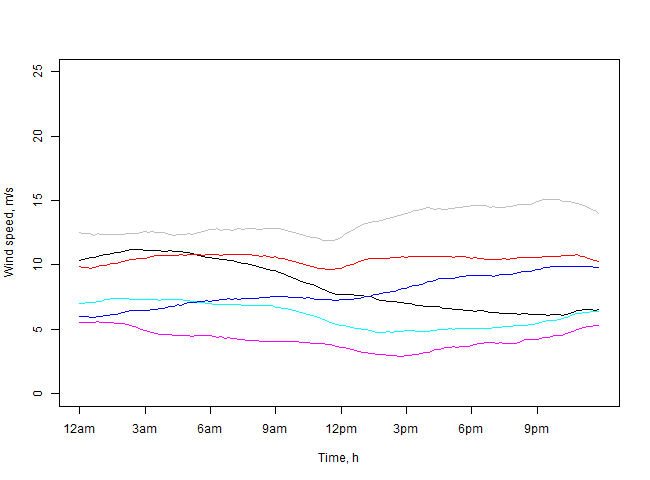}
\caption{The representatives obtained by applying k-means with six clusters to the dataset. Mean behavior is dominant.}
\label{kmeans}
\end{figure}

Much work has been done on the choice of distance functions in high-dimensional spaces. The intuitive choice, and a typical method, is to extend Euclidean spatial distance to higher dimensions using the $L_p$ norm: for two $n$-dimensional observations $\mathbf{x}$ and $\mathbf{y}$, $L_p(\mathbf{x},\mathbf{y}) = \big( \sum_{i=1}^n |x_i - y_i|^p \big) ^ {1/p}$. For example, $p=2$ leads to the the familiar RMSE (root mean squared error). This approach, however, has several drawbacks. First, it is sensitive to small differences in level between observations, emphasizing the mean behavior. For example, applying k-means clustering to the dataset yields remarkably uninformative groups, and has the additional disadvantage that the cluster centers, as pointwise means of all cluster members, are unrealistically smooth (see Figure \ref{kmeans}). Indeed, \cite{kusiak} generate clusters based on various parameters using k-means (in their case, in order to fit a different short-term predictive model of power generation for each cluster), and find that performing this clustering with wind speed does not lead to better models.

The $L_p$ norm is also dominated by observations' behavior on dimensions, or in our case at times, where differences are large; if there is heteroskedasticity across times, those times with higher variation will tend to contribute the most to the distance calculation. \cite{beyer} describe the problem of loss of contrast as the number of dimensions grows (that is, the distance from an observation to its nearest neighbor becomes, relatively, about the same as the distance to its farthest neighbor), and \cite{aggarwal_Lp} demonstrate that the Euclidean norm is particularly subject to this problem, noting that this sensitivity to heteroskedasticity may be to blame. In a similar vein, the $L_p$ norm relies on absolute distance between elements of $\mathbf{x}$ and $\mathbf{y}$, and may thus also be sensitive to skew in the distribution of observations; and, more broadly, it does not consider the observations in the context of the rest of the dataset. While we might attempt to adjust for heteroskedasticity and skew by assigning weights to the dimensions in the distance calculation, or applying transformations to the observations themselves, these approaches would require expert choices of functions and parameters. 

\cite{aggarwal_IGrid} propose a method called \emph{IGrid}-index based on a similarity score called $PIDist$, which manages the problem of heteroskedasticity across dimensions by calculating equidepth ranges on each dimension; $\mathbf{x}$ and $\mathbf{y}$ are assigned a similarity based on (i) the number of dimensions on which they fall into the same range, and (ii) their absolute proximity on each of these dimensions, as a proportion of the range. The similarities, and distinctions, between the $PIDist$ function and the metric proposed here will be discussed further in section 2.

To emphasize differences in shape, or to define the similarity of pairs of observations in a way that takes the rest of the dataset into account, a different approach is needed. Note that these methods could be of use in other applications where the shape of time series or functional data is a concern, such as tracking meteorological data, growth curves, or the number of users of a system over time. \cite{feng} explore selecting scenarios for the time series of future energy demand and fuel cost, in the context of generation expansion planning (GEP); while they use a multistage selection method instead of single-stage clustering, their method still relies on finding the pairwise distances between scenarios.

We turn to the methods of depth statistics, which are designed to express the centrality of observations relative to a dataset, without reference to external scales or parameters. In the one-dimensional case, the depth statistic corresponds to the median and quantiles of a dataset, with the median being the deepest, or most central, point. This concept can be extended to different types of data. For example, \cite{liu90} gives a version, the \emph{simplicial depth}, for the centrality of points in $\Re^p$ relative to a distribution in $\Re^p$. This depth measures how often the observation of interest falls inside a simplex determined by $p+1$ points drawn from the distribution; those points which fall inside such random simplices more often are considered more central.

Another extension is the \emph{band depth}, developed by \cite{lpr09} to judge typical or atypical shapes of functional observations. Instead of simplices, Lopez-Pintado and Romo use bands defined by two or more observations drawn from a fixed dataset. At each point in the domain, the upper limit of the band is the maximum value of all the observations defining the band, while the lower limit is the minimum value of all these observations. In the original band depth, each observation is compared to each band; if the observation falls within the band's limits at each point in the domain it is considered to lie within the band. Those observations falling within the most bands are considered the most central. 

With many datasets, however, observations successfully lie within very few bands, causing many ties in depth between multiple observations. Lopez-Pintado and Romo thus introduce the \emph{generalized band depth}. Here, instead of receiving a binary ``in or out" score for each band, an observation receives a score for each band corresponding to the proportion of the domain for which the observation falls within the band's limits. This version greatly reduces ties, and it allows us to use only two observations to define each band (while in the original version, bands defined using only two observations often failed to contain \emph{any} other observations, requiring the use of more forgiving, but less informative and more computationally expensive, three-observation bands).

The band depth can easily be adapted for use in classification of data; for example, given two preexisting groups and a new observation to be classified, we could simply place the observation in the group where it would be more central. In itself, however, it cannot be used for unsupervised clustering. Since we wish to generate representatives and clusters from our dataset without prior knowledge, we extend the concept of the generalized band depth to a distance measure that yields pairwise distances between observations, allowing the use of any clustering method based on such pairwise distances.

Our methodology is described in Section \ref{sec:method}. In Section \ref{sec:sims} we apply the method to simulated data, and in Section \ref{sec:wind_res} we examine the application to wind speed data, including transformations of the data obtained by removing typical seasonal behavior and using a time-frequency representation, and obtain clustering results in each case. Section \ref{sec:conclu} describes our conclusions and future research. Finally, the Appendix gives a proof that our method defines a valid distance metric.

%%%%%%%%%%%%%%%%
%%%% METHODOLOGY %%%%

\section{Methodology}
\label{sec:method}

Following the generalized band depth developed in \cite{lpr09}, a band $\mathbf{b}=({b}_{\ell},{b}_u)$ is defined by a subset of the observations; in our work, we use bands defined by two observations, and denote the set of all such bands by $\mathbf{B}$. In the analysis below we assume a domain of discrete time points, though the method may be extended to more general domains by using an appropriate measure. (Of course, the domain need not be time at all, nor even have an ordering on the dimensions.) Thus, at each time point $t$, the upper and lower edges of the band are defined by the region bounded by the two observations. Hence a band $\mathbf{b}$ defined by the observations $v$ and $w$ has
\begin{eqnarray*}
{b}_{\ell}(t|v,w) &= \min(v(t),w(t)); \hspace{5mm}
{b}_{u }(t|v,w) =& \max(v(t),w(t)).
\end{eqnarray*}
An observation $x$ lies within the band $\mathbf{b}$ (or $\mathbf{b}$ contains $x$) at index $t$ if $x(t)$ is in $[b_{\ell}(t), b_u (t)]$. We use $I^\mathbf{b}[x(t)]$ to denote an indicator function
\[ I^\mathbf{b}[x(t)] = \begin{cases} 1 &\mbox{if } {b}_{\ell}(t) \le x(t) \mbox{ and } x(t)\le {b}_{u}(t) \\ 
0 & \mbox{otherwise}. \end{cases} \]
For any band and observation, there is a set of time points, possibly empty, at which the observation falls within the band, which we denote $T^\mathbf{b}(x)=\{ t:I^\mathbf{b}[x(t)]=1\}$. The sizes of such sets can be used to obtain a measure of centrality for the observation $x$. We go further, however, to define a similarity score for any given band by adapting the Jaccard similarity. For two observations $x$ and $y$ and a band $\mathbf{b}$, the bandwise similarity is defined as
\[
s^{\mathbf{b}}_{xy} = \frac{|T^{\mathbf{b}}(x) \cap T^{\mathbf{b}}(y)|}{|T^{\mathbf{b}}(x) \cup T^{\mathbf{b}}(y)|}.
\]
Following the original Jaccard similarity, we define this quantity to be 1 when the denominator is zero (that is, for bands which never contain either $x$ or $y$). Again following the Jaccard similarity metric, the bandwise distance is defined by subtracting the bandwise similarity from 1, so that $d^{\mathbf{b}}_{xy} = 1 - s^{\mathbf{b}}_{xy}$.
We then define the overall similarity between $x$ and $y$ to be the average of all bandwise similarities for ``informative" bands. Specifically, let $B_{xy}$ be the set of all bands that contain either $x$ or $y$ at any index, $B_{xy} = \{ \mathbf{b} : \sum_t I^\mathbf{b}[x(t)] + \sum_t I^\mathbf{b}[y(t)] > 0 \}$. We obtain the average of the bandwise similarity scores only over bands in this set, and call this the overall similarity,
\[ S_{xy} = \frac{1}{|B_{xy}|} \sum_{\mathbf{b} \in B_{xy}} s^{\mathbf{b}}_{xy}. \]
Likewise, the band distance between $x$ and $y$, $D_{xy}$, is defined as
\[ D_{xy} = \frac{1}{|B_{xy}|} \sum_{\mathbf{b} \in B_{xy}} d^{\mathbf{b}}_{xy},\]
which is equivalent to $D_{xy} = 1 - S_{xy}$.

We also note that $D_{xy}$ has several desirable characteristics as a measure of distance.
\begin{thm}
$D_{xy}$ is a distance metric: that is, for any $x$, $y$, and $z$, we have $D_{xy} \ge 0$ (non-negativity); $D_{xy} = 0$ if and only if $x=y$ (identity of indiscernibles); $D_{xy}=D_{yx}$ (symmetry); and $D_{xz} \le D_{xy} + D_{yz}$ (triangle inequality).
\end{thm}
A complete proof is found in the Appendix. As is often the case, non-negativity and symmetry are trivial, and identity of indiscernibles is simple to show. The proof that $D_{xy}$ fulfills the triangle inequality is considerably more involved. Given a band $\mathbf{b}$, the bandwise distance $d^{\mathbf{b}}_{xy}$ fulfills the triangle inequality because it is a Jaccard distance; an average of such distances would likewise be a distance metric. But to obtain $D_{xy}$ we take the average of these scores only over the set $B_{xy}$ of bands that contain either $x$ or $y$ at some time, and the sets $B_{xy}$, $B_{yz}$, and $B_{xz}$ are not necessarily the same, nor the same size.

We could consider these $d^\mathbf{b}_{xy}$ sums as being over all possible bands, rather than only those in $B_{xy}$, since the contribution to the sum $\sum_{\mathbf{b}\in \mathbf{B}} d^{\mathbf{b}}_{xy}$ is zero for bands not in $B_{xy}$; but note that we divide only by the size of $B_{xy}$, not $|\mathbf{B}|$, the size of the entire set of bands. While dividing by either quantity serves to produce distances that range conveniently between 0 and 1, there are several reasons for using $|B_{xy}|$. Primarily, this approach means that bands which never contain either observation have no effect on the distance between the observations (since these bands contribute to neither $|B_{xy}|$ nor $\sum_{B_{xy}} d^{\mathbf{b}}_{xy}$), and can be considered as providing no information about the observations' similarity. Were we to divide by the total number of bands $|\mathbf{B}|$, each band that contained neither $x$ nor $y$ would serve to decrease $D_{xy}$, since $d^\mathbf{b}_{xy}$ for that band would be 0 while the band still contributed to $|\mathbf{B}|$. Thus the introduction of new bands into the dataset would affect $D_{xy}$ even if those bands never contained either $x$ or $y$. (Note that, with the present method of defining bands, the introduction of any new \emph{observation} into the dataset still affects $D_{xy}$ through, at a minimum, those bands defined by the new observation and $x$ or $y$. So the entirety of the dataset is taken into account in any event.)

Furthermore, since $d^\mathbf{b}_{xy}$ on a band that \emph{does} contain $x$ or $y$ is not zero unless $x$ and $y$ are in the band at exactly the same time points, a band containing neither observation would add less to the overall distance than a band containing $x$ and $y$ at nearly the same times. Yet the latter situation clearly gives more evidence for the similarity of $x$ and $y$ than the former.

The resulting band distance has several useful properties. It is entirely data-driven, describing the distance between $x$ and $y$ in the context of the other observations in the dataset. It is invariant under any order-preserving operation on the observations in the dataset; we may apply monotonic transformations to all observations, including monotonic transformations which are not consistent across time. And, as we shall see in the next section, it treats all time points equally: those times where variation between observations is (in absolute terms) large do not dominate the overall calculation of distances.

It is interesting to note the distinctions between the band distance and the $PIDist$ method of calculating high-dimensional similarity proposed by \cite{aggarwal_IGrid}. The $PIDist$ similarity is, like the band distance, explicitly designed to take the overall dataset into account when looking at the similarity of observations. Like the $L_p$ norm, it is a family of similarity functions indexed by $p$, and is calculated as follows. On each dimension $t$, divide the values of all the observations in the dataset into $k$ equidepth groups (that is, groups which contain equal numbers of points). The highest and lowest observation values in each group generate the \emph{ranges}. For two observations of interest ${x}$ and ${y}$, there is some set $S({x},{y},k)$ of dimensions on which $x(t)$ and $y(t)$ are in the same range. For each such dimension, let $r_t$ be the width of the range into which $x(t)$ and $y(t)$ fall. Then the overall similarity of ${x}$ and ${y}$ is

\[ PIDist({x},{y}|k,p) = \left[ \sum_{t \in S({x},{y},k)} \left(1 - \frac{|x(t) - y(t)|}{r_t} \right)^p \right] ^{1/p}. \]

Note that \emph{PIDist} defines a similarity, and must be converted to a distance and, if desired, rescaled to give values between 0 and 1. It is also not shown to be a distance metric. \footnote{Many distance functions do not fulfill the necessary conditions to be a metric, but there is utility as well as theoretical appeal in doing so: for example, \cite{gunopulos} note that many retrieval algorithms for finding data points similar to a given observation require the triangle inequality.}

The \emph{PIDist} method is distinct from the band distance since the observations that define bands are not necessarily those that define ranges; but there are also conceptual differences. While the use of equidepth ranges 
helps deal with differing levels of variation across dimensions, on each dimension \emph{PIDist} still uses the absolute distance between the observations, $|x(t) - y(t)|$. As a result, the method is not driven solely by the relative values of the observations in the dataset; it may remain sensitive to skew, and it is not invariant under order-preserving transformations as is the band distance. In addition, \emph{PIDist} examines each dimension separately, whereas the band distance looks at ${x}$ and ${y}$ relative to the behavior of other observations across all dimensions. Finally, the number of ranges $k$ must be chosen by the user; Aggarwal notes that the optimal value appears to vary with the dimension of the data, and provides some suggestions for its selection, but it remains an externally set parameter.

%%%%%%%%%%%%%%%
%%%% SIMULATIONS %%%%

\section{Simulation study}
\label{sec:sims}

To assess the performance of our metric, we use simulated time series data. To compare the performance of the different metrics, we generate a reference set containing observations of known types; then for each metric, we create a pairwise distance matrix, perform clustering with a fixed number of clusters, and calculate the Rand index (introduced in \cite{rand}) between the resulting classification and the true classification. The difference between the Rand index values for the band distance and for Euclidean distance, $\Delta_R$, indicates which method gave a more accurate clustering, with positive values counting as a win for the band distance. By repeating this process over $M$ multiple runs, we can obtain mean and standard deviation for $\Delta_R$, as well as a standardized z-score, $Z_R = \bar{\Delta_R} /\sqrt{(\widehat{Var}(\Delta_R)/{M})}$.

Given that we can use any clustering method that relies on pairwise distances between observations, for the current analysis we use k-medoids. This method is similar to k-means, in that it is an iterative procedure; at each iteration, observations are assigned to clusters based on which cluster center is closest, then cluster centers are updated based on the new cluster memberships. In k-medoids, the cluster centers are always observations from the original dataset, specifically the most central observation in each cluster (by contrast, in k-means, the cluster centers are the pointwise means of the cluster members). K-medoids accepts a pairwise distance matrix as input, and it yields sensible representatives for each cluster rather than the overly smoothed aggregate output of k-means. The R package ``cluster" (see \cite{R-cluster}) provides a tool for performing k-medoids clustering, which automatically selects an initial set of medoids based on the dataset.

%%%%SIMULATION 2MU%%%%

\paragraph{Simulation A.} Since we are interested in examples of data with nonconstant variance, we simulate a set of observations from a multivariate normal distribution displaying heteroskedasticity. In a simple case, we have two classes of observations, each of length 15, with the mean vectors
\[
[\boldsymbol{\mu}_1(t)] = (0.1, 0.2, 0.3, \dots, 1.4, 1.5)', \hspace{2mm}
[\boldsymbol{\mu}_2(t)] = (1.5, 1.4, 1.3, \dots 0.2, 0.1)'.
\]

\begin{figure}[H]
\centering
\includegraphics[width=\halfwidthcoef \textwidth]{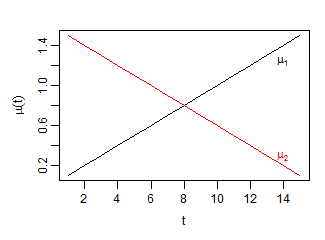} 
\includegraphics[width=\halfwidthcoef \textwidth]{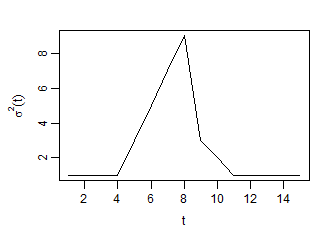}
\caption{Simulation A: mean vectors for each class, $\boldsymbol{\mu}_1(t)$ and $\boldsymbol{\mu}_2(t)$, and variance vector, $\boldsymbol{\sigma}^2(t)$.}
\label{sim2MU_params}
\end{figure}

We designate a nonconstant variance vector over time, constant over both classes for simplicity,
\[[\boldsymbol{\sigma}^2(t)] =  (1, 1, 1, 1, 3, 5, 7, 9, 3, 2, 1, 1, 1, 1, 1)' \]
and a coefficient $\rho = 0.9$ to represent autocorrelation. Then, letting $\boldsymbol{\rho} =  (\rho^0, \rho^1, \rho^2, \dots, \rho^{14})'$, for both classes the covariance matrix is
\[
\Sigma = \boldsymbol{\sigma}' * Toep(\boldsymbol{\rho}) * \boldsymbol{\sigma},
\]
where $Toep(\boldsymbol{\rho})$ is the Toeplitz matrix formed using the vector $\boldsymbol{\rho}$. The mean and variance functions are shown in Figure \ref{sim2MU_params}.

We simulate 10 observations of each type, and compare the results of clustering these curves with the known classification based on the $\boldsymbol{\mu}$ vector used to generate each. As described, we generate two pairwise distance matrices, one using Euclidean distance and one using band distance; then, in each case, we cluster using k-medoids with a fixed number of clusters. We perform the simulation $M=1000$ times. The average value of $\Delta_R$ is positive, and the standardized $Z_R$ value is 9.58, indicating that the band distance significantly outperforms Euclidean distance.

%%%%SIMULATION 3MU%%%%

\begin{figure}[H]
\centering
\includegraphics[width=\halfwidthcoef \textwidth]{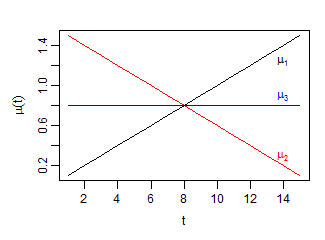}
\includegraphics[width=\halfwidthcoef \textwidth]{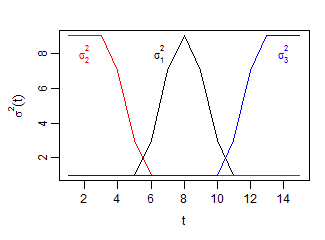}
\caption{Simulation B: mean and variance vectors. Each combination of $\boldsymbol{\mu}$ and $\boldsymbol{\sigma}^2$ defines a class.}
\label{sim3MU_params}
\end{figure}

\paragraph{Simulation B.} For a slightly more complicated test, we consider observations generated from nine different heteroskedastic multivariate normal distributions. The structure is as described above, with $\rho=0.9$, but we have three different $\boldsymbol{\mu}$ mean functions and three different $\boldsymbol{\sigma}^2$ variance functions, shown in Figure \ref{sim3MU_params}, for nine possible combinations. 

As before, we generate 10 observations of each type and run $M=1000$ simulations; the band distance is again superior, with an even higher $Z_R$ value of 23.9.

%%%%SIMULATION PGRAMS%%%%

\begin{figure}[H]
\centering
\includegraphics[width=\halfwidthcoef \textwidth]{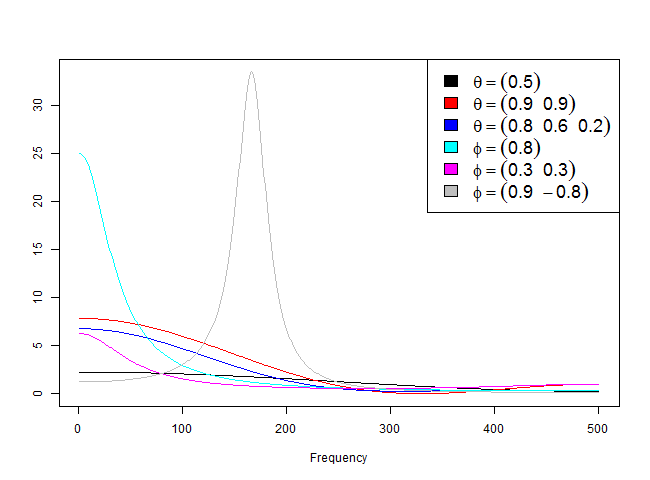}
\includegraphics[width=\halfwidthcoef \textwidth]{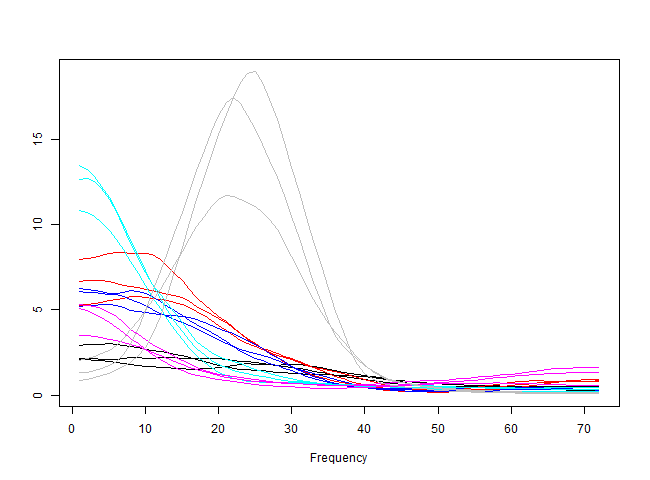}
\caption{Simulation C: spectral densities of each model (left); sample of smoothed periodograms of simulated time series, showing three observations of each type (right).}
\label{simPGRAMS_spec-sam}
\end{figure}

\begin{figure}[H]
\centering
\includegraphics[width=\widthcoef \textwidth]{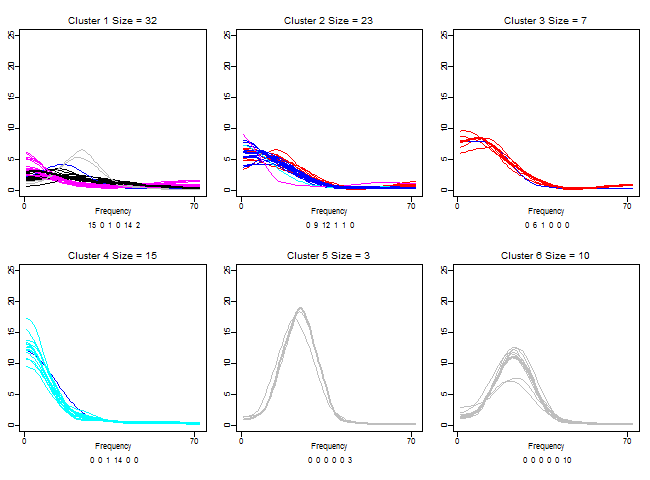}
\caption{Results of k-medoids clustering using Euclidean distance. Numbers of observations of each type are given below each cluster.}
\label{simPGRAMS_eucl}
\end{figure}

\begin{figure}[H]
\centering
\includegraphics[width=\widthcoef \textwidth]{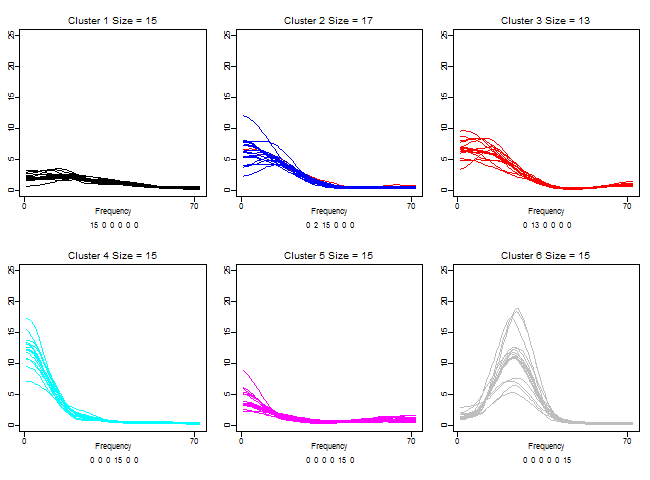}
\caption{Results of k-medoids clustering using band distance. Numbers of observations of each type are given below each cluster.}
\label{simPGRAMS_dMBD}
\end{figure}

\paragraph{Simulation C.} To see how this behavior plays out in less artificial data, we simulate time series from six different stationary ARMA models, and examine their periodograms. The six models are as follows: MA(1), $\boldsymbol\theta = 0.5$; MA(2), $\boldsymbol\theta = (0.9, 0.9)$; MA(3), $\boldsymbol\theta = (0.8, 0.6, .2)$; AR(1), $\boldsymbol\phi = 0.8$; AR(2), $\boldsymbol\phi = (0.3, 0.3)$; AR(2), $\boldsymbol\phi = (0.9, -0.8)$, all with error variance equal to $1.0$. We generate 15 realizations from each model, for a total sample of 90 simulated time series, each with 144 time points for comparability to the wind data. We then obtain the smoothed periodogram of each realization, using a modified Daniell kernel; we need not take the short-time Fourier transform here since the underlying time series are stationary. The spectral density of each model, and a sample of these periodograms, are shown in Figure \ref{simPGRAMS_spec-sam}. The clustering results from one simulation run are shown in Figures \ref{simPGRAMS_eucl} and \ref{simPGRAMS_dMBD} respectively.

Here, Euclidean distance commits several errors, subdividing one class and combining others, with a Rand index of 0.838 when compared to ground truth. Clustering based on the band distance, however, gives substantially more correct assignments, misclassifying only two observations (for a Rand index of 0.986). With $M=1000$ runs, we obtain a $Z_R$ value of 101.04, a convincing argument that the superior performance of the band distance shown above is not anomalous.

We can ascribe the errors made with Euclidean distance in large part to the heteroskedasticity of the data. The periodograms do not have equal variance at each frequency (the variance increases with the amplitude). The resulting large discrepancies between observations, even observations of the same class, at these frequencies dominate the Euclidean calculation of distance. In the band distance, which relies only on the ordering of observations at each point rather than their difference on an absolute scale, these frequencies receive no more ``weight" than those at which overall variance is lower.

%%%%%%%%%%%%%%
%%%% WIND DATA %%%%

\section{Application: wind data}
\label{sec:wind_res}

The proposed distance metric can be applied to any high-dimensional real-valued observations. We can, of course, apply it directly to the daily wind speed curves; but there are also certain transformations that may make the data more informative.

\subsection{Seasonal behavior}

\begin{figure}[H]
\centering
\includegraphics[width=\halfwidthcoef \textwidth]{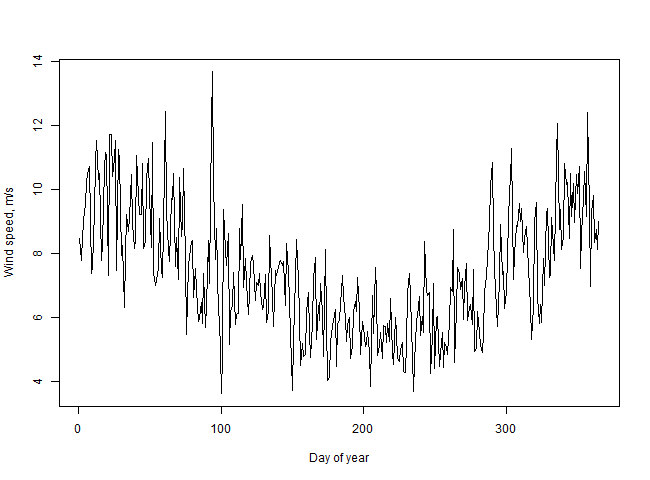}
\includegraphics[width=\halfwidthcoef \textwidth]{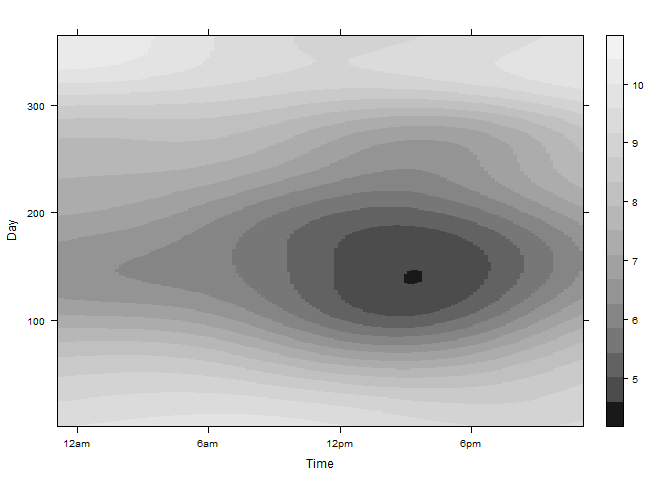}
\caption{Left, mean wind speed by day, over 3 years (2004-2006). Right, typical intra-day wind speed curves for each day of the year.}
\label{daily_mean-GAM}
\end{figure}

Although wind behavior is highly variable from day to day, there are seasonal trends. For example, Figure \ref{daily_mean-GAM} shows that average wind speeds each day are higher in winter. There is also a typical (though loosely followed) shape to the intra-day wind speed curve, which changes over the course of the year. By identifying this shape and removing it from our observations, we can emphasize departures from typical behavior. To this end, we fit a generalized additive model of the form $W = s(t,c)$, where $W$ represents wind speed, and $s(t,c)$ is some smooth function of time of day, $t$, and calendar day of year, $c$. The model requires that the expected wind speed change smoothly over the course of the day, and also that the daily curve change smoothly over the course of the year. For the purposes of fitting this model we have three replicates of each time-day combination, corresponding to the three years of observations. We fit the model using the ``gam" function from the R package ``mgcv" (see \cite{R-mgcv}), which uses Newton's method. Figure \ref{daily_mean-GAM} shows the result. The typical day has higher wind speed at night, with a low point in the afternoon; this pattern is more pronounced in the summer, though overall levels are higher in winter. 

\begin{figure}[H]
\centering
\includegraphics[width=\widthcoef \textwidth]{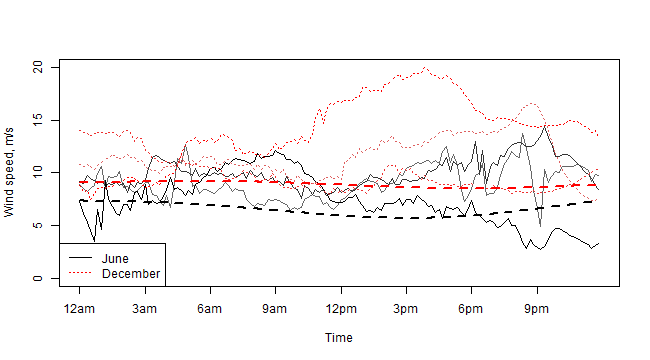}
\includegraphics[width=\widthcoef \textwidth]{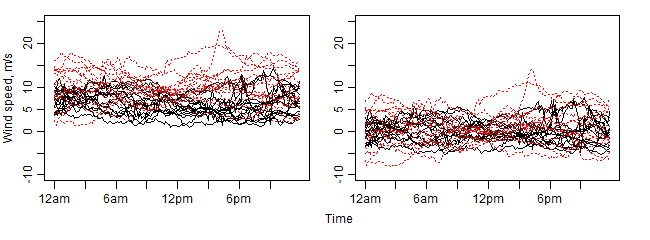}
\caption{Top, observations from June (solid lines) and December (dotted lines), with the typical daily curves shown as dashed lines. Bottom, the first 15 days of June (solid) and December (dotted) in year 1. Left, orignal observations; right, typical daily curves removed.}
\label{DJ90_sam-cpar}
\end{figure}

We can then remove this typical daily curve from our observations. Figure \ref{DJ90_sam-cpar} shows these typical daily curves as compared to the original wind speed observations for several days in June and December. We also see a comparison of original observations with the corresponding residual curves formed by removing the typical daily curve. The predominant effect is to shift the center of the observations; thus, removing the typical daily curve is valuable where we are concerned more with individual days' shape than with the slow seasonal change in level.

\subsection{Time-frequency analysis}

\begin{figure}[H]
\centering
\includegraphics[width=.8\textwidth]{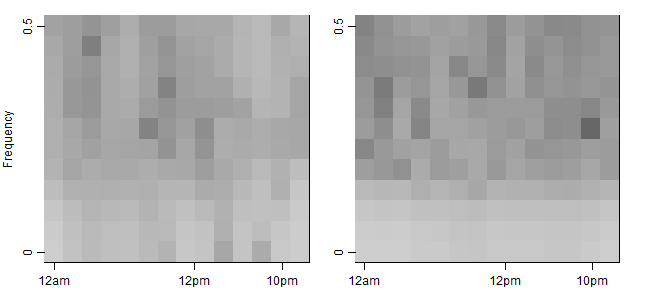}
\includegraphics[width=.8\textwidth]{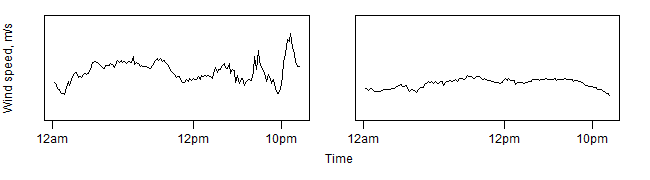}
\caption{Log-scale STFTs, above original time series from which they are calculated. Time windows proceed from left to right, frequencies from bottom (lowest frequency) to top (highest frequency). Loadings proceed from dark (low loading) to light gray (high loading).}
\label{DJ90_stft_sam}
\end{figure}

We may also wish to emphasize short-term variability in wind speeds. In our application, sharp drops and peaks in wind speed are of particular importance: for example, a day when wind speed increases slowly requires different decisions in the power system than a day when wind speed increases erratically, with sharp climbs and drops. To this end, we can look at wind behavior in the frequency domain, by taking the Fourier transform of the daily time series and examining days' loadings on different frequencies. In this way, days with similar short-term erratic behavior will be marked as similar; smoothly varying days, with high power only on low frequencies, will likewise be similar to one another. Crucially, however, the wind speed series are not stationary over the course of the day; thus we use a time-frequency approach. By examining the short-time Fourier transforms of the wind speed series, we can see how frequency behavior changes across overlapping time windows. We use the R function ``stft" from package ``e1071" (see \cite{R-e1071}). In Figure \ref{DJ90_stft_sam} we see two example STFTs on the log scale, one from a day with erratic changes in wind speed and one from a day with smooth changes. In this case we have applied the STFT to data with the typical daily wind curve removed.

Note that the more erratic day, on the left, has significant scores on nearly all frequencies, particularly in the evening. In contrast, the day with smooth wind changes has almost no loading on the higher frequencies.

By vectorizing these images, we obtain high-dimensional data to which we can apply our distance measure. Note that this transformation loses some information about the original series, since the limited number of rows in the STFT means we truncate the set of frequencies for which we have loadings. In addition, the overall level of the series is lost; here, our inputs were detrended series, but depending on the application, it may be advisable to include the mean as additional dimensions of the data. We may also choose to normalize by the overall variance within each day (perhaps including this value, too, as another element), so that the STFTs all have the same mean value. Then we are left with vectors that express only local variation.

\subsection{Clustering results}

\begin{figure}[H]
\centering
\includegraphics[width=\widthcoef \textwidth]{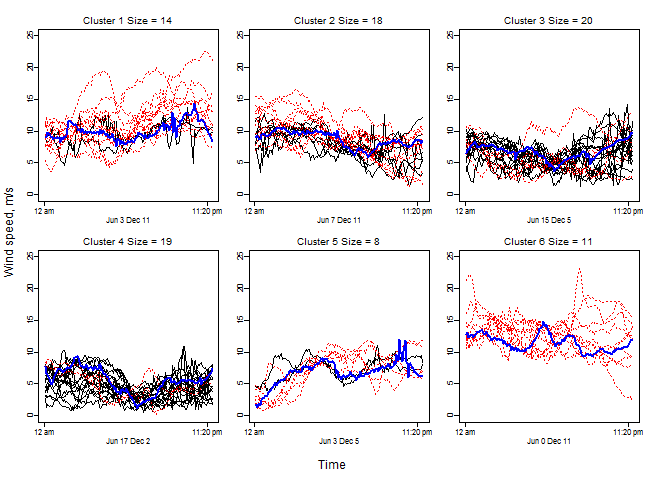}
\caption{Clustering results for original wind speed data, with number of June (solid) and December (dotted) observations in each cluster. Cluster centers (medoids) shown in bold.}
\label{DJ90_orig}
\end{figure}

We examine a sample of 90 observations, corresponding to the first 15 days of June and December in each year of the dataset. As above, we use k-medoids clustering and specify a fixed number of clusters; here we use six, corresponding to a common desired number of representative days in our application. First, we apply the distance metric and clustering directly to the wind speed time series, with results shown in Figure \ref{DJ90_orig}.

\begin{figure}[H]
\centering
\includegraphics[width=\widthcoef \textwidth]{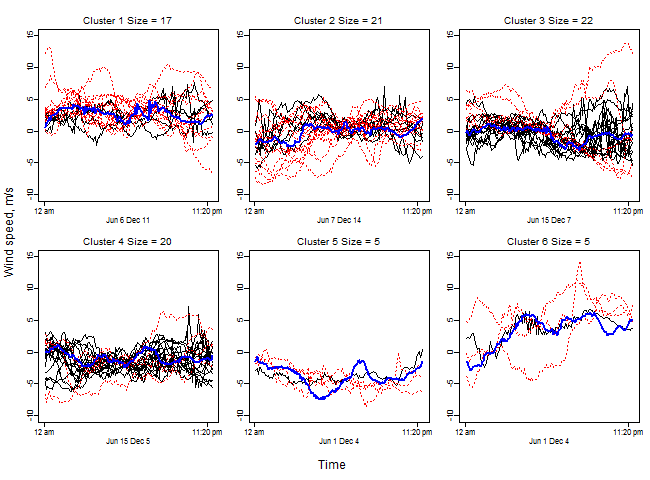}
\caption{Clustering results for observations with typical daily shape removed, with number of June (solid) and December (dotted) observations in each cluster. Cluster centers (medoids) shown in bold.}
\label{DJ90_dmGAM}
\end{figure}

We can see that in many cases observations from the same month are clustered together, reflecting the different average levels in winter and summer. Using the corresponding series with the GAM-based intra-day trend removed yields notably different results (the Adjusted Rand Index between the two classifications of the observations is only 0.126). In the latter case, shown in Figure \ref{DJ90_dmGAM}, clusters tend to contain a more even mix of summer and winter days.

\begin{figure}[H]
\centering
\includegraphics[width=\widthcoef \textwidth]{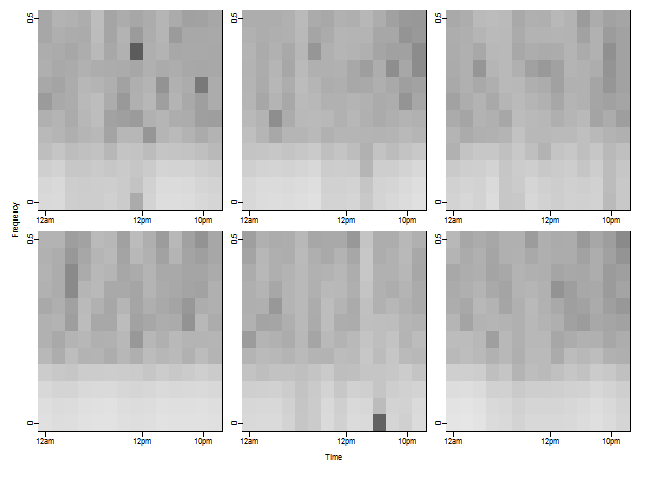}
\caption{Medoids obtained by clustering on STFTs of observations with typical daily shape removed. Time windows proceed from left to right, frequencies from bottom (lowest frequency) to top (highest frequency). Loadings are on the log scale, from dark (low loading) to light gray (high loading).}
\label{DJ90_stft_centers}
\end{figure}

Finally, we can obtain the STFTs of these detrended days, to put maximum emphasis on short-term variation. Even with a low resolution, using only 14 overlapping time windows and 12 Fourier coefficients, the representative STFTs for each cluster show distinctly different behaviors (see Figure \ref{DJ90_stft_centers}). For example, cluster 5 incorporates days with high-frequency behavior in the mid-afternoon, indicating sharply varying speeds at that time of day; in cluster 6, meanwhile, we see a tendency toward middle-frequency behavior at the beginning of the day.

\begin{figure}[H]
\centering
\includegraphics[width=\widthcoef \textwidth]{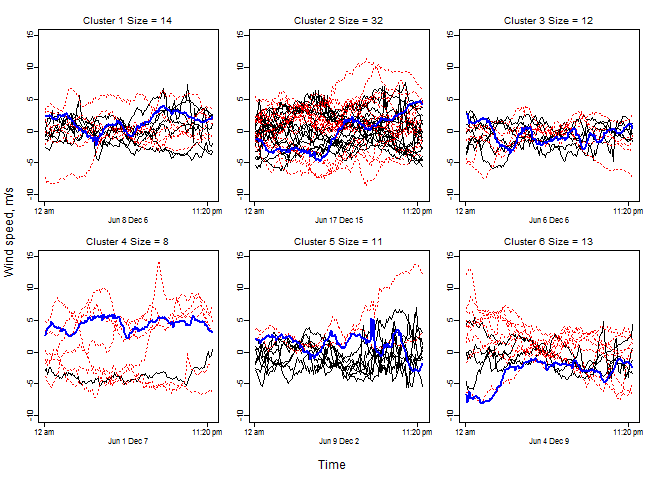}
\caption{Clustering results for STFTs of observations with typical daily shape removed, with number of June (solid) and December (dotted) observations in each cluster. Cluster centers (medoids) shown in bold.}
\label{DJ90_stft_ws}
\end{figure}

By looking at the original data for the days falling into each cluster, shown in Figure \ref{DJ90_stft_ws}, we can see how these STFTs correspond to the original wind speed curves. While the observations are noisy, we can still observe the time-frequency behaviors discussed above. Note that overall level information is lost by using the STFTs; the emphasis here is on the changes in variability over time, as might be desired for making decisions about what power sources must be available to back up contributions from wind.

%%%%%%%%%%%%%%%
%%%% CONCLUSION %%%%

\section{Conclusion}
\label{sec:conclu}

In this paper, we have introduced a distance metric based on the idea of the depth statistic (in particular, the generalized band depth of \cite{lpr09}), combined with the Jaccard distance measure. Our metric provides pairwise distances between high-dimensional observations, avoiding reliance on pointwise Euclidean distance and taking into account the behavior of the entire dataset of interest. Using the resulting pairwise distance matrix, we are able to perform clustering using k-medoids.

We applied our methods to wind speed time series, drawn from the EWITS dataset, to obtain clusters of days demonstrating different wind behaviors as well as representative days of each type. To examine departures from typical behavior, we removed a smoothly varying typical daily curve from the observations; additionally, we used the short-term Fourier transforms of each day to obtain a time-frequency representation that emphasized local variation in wind speed.

In the future, we will examine additional ways of extracting information from the raw wind speed curves, in particular ways of reducing their dimension for more effective clustering. For example, we can use use a constrained factor model for the time-frequency representations of the data, in which each day's STFT is modeled using empirical orthogonal functions (constrained to be consistent across days) with different loadings. We can also seek covariates or recognizable meteorological characteristics associated with each cluster to increase interpretability.

Finally, although we have focused on wind speed at a single site in this analysis, it is important to remember that a spatial component exists and is non-trivial. There is strong association between wind speeds at the same site over short timeframes, and weaker correlation between wind speeds at different sites at approximately the same time, with a possible lag dependence in the prevailing wind direction. This inter-site correlation suggests possibilities for handling clustering across multiple locations in a more sophisticated manner than simply concatenating individual sites' time series; but the nature of the correlation changes depending on the particular sites and on the time of day, and future work must take this variation into account. 

%%%%%%%%%%%%%
%%%% APPENDIX %%%%

\section{Appendix}
\label{sec:proof}

Consider a set of observations $X$ consisting of at least three observations, and select three arbitrary representatives $x$, $y$, and $z$ from this set. We prove that $D$ is a distance metric. 

\begin{longtable}{p{.10\textwidth} p{.90\textwidth} }

$\mathbf{Notation}$ & \\
$t$ & time index \\
$X$ & set of observations \\
$x,y,z$ & observations \\
$x(t)$ & observation at time $t$ \\
$N$ & number of observations, $|X|$ \\
$\mathbf{b}$ & band \\
$\mathbf{B}$ & set of all bands \\
$|\mathbf{B}|$ & number of bands, $N \choose{2}$ \\
$b_{u}(t)$ & upper limit of band $\mathbf{b}$ at time $t$ \\
$b_{\ell}(t)$ & lower limit of band $\mathbf{b}$ at time $t$ \\
$I^\mathbf{b} [x(t)]$ & indicator that observation $x$ is in band $\mathbf{b}$ at time $t$: \\
 & \hspace{4mm} $I^{\mathbf{b}}[x(t)] \mbox{ if } x(t) \le b_{u}(t)$ and $x(t) \ge b_{\ell}(t)$, else $I^{\mathbf{b}}[x(t)]=0$ \\
$T^{\mathbf{b}}(x)$ & set of times when observation $x$ is in band $\mathbf{b}$: \\
 & \hspace{4mm} $T^{\mathbf{b}}(x) = \{ t : I^{\mathbf{b}}[x(t)] = 1 \}$ \\
$B_{xy}$ & set of bands into which either $x$ or $y$ passes at any time: \\
 & \hspace{4mm} $B_{xy} = \{ \mathbf{b} : \sum_t I^{\mathbf{b}}[x(t)] + \sum_t I^{\mathbf{b}}[y(t)] > 0 \}$ \\
$s^{\mathbf{b}}_{xy}$ & Jaccard similarity between $x$ and $y$ for band $\mathbf{b}$: \\
 & \hspace{4mm} $s^{\mathbf{b}}_{xy} = \frac{|T^{\mathbf{b}}(x) \cap T^{\mathbf{b}}(y)|}{|T^{\mathbf{b}}(x) \cup T^{\mathbf{b}}(y)|} $ if ${|T^{\mathbf{b}}(x) \cup T^{\mathbf{b}}(y)|} \ne 0$,  \\
 & \hspace{4mm} else $s^{\mathbf{b}}_{xy} = 1$ \\
$d^{\mathbf{b}}_{xy}$ & Jaccard distance between $x$ and $y$ for band $\mathbf{b}$: \\
 & \hspace{4mm} $d^{\mathbf{b}}_{xy} = 1 - s^{\mathbf{b}}_{xy}$ \\
$S_{xy}$ & overall similarity score for $x$ and $y$: \\
 & \hspace{4mm} $S_{xy} = \frac{1}{|B_{xy}|} \sum_{B_{xy}} s^{\mathbf{b}}_{xy}$ \\
 & \hspace{4mm} (average Jaccard similarity between $x$ and $y$ \\
 & \hspace{4mm}  over all bands containing $x$ or $y$ at any time) \\
$D_{xy}$ & overall distance between $x$ and $y$: \\
 & \hspace{4mm} $D_{xy} = 1 - S_{xy}$ \\
 & \hspace{4mm} or $D_{xy} = \frac{1}{|B_{xy}|} \sum_{B_{xy}} d^{\mathbf{b}}_{xy}$

\end{longtable}

\subparagraph{Non-negativity}

We must show $D_{xy} \ge 0$. To this end, note that for any $\mathbf{b}$, $|T^{\mathbf{b}}(x) \cap T^{\mathbf{b}}(y)| \le |T^{\mathbf{b}}(x) \cup T^{\mathbf{b}}(y)|$; thus, $s^{\mathbf{b}}_{xy} = \frac{|T^{\mathbf{b}}(x) \cap T^{\mathbf{b}}(y)|}{|T^{\mathbf{b}}(x) \cup T^{\mathbf{b}}(y)|} $ is necessarily less than or equal to 1, and so $d^{\mathbf{b}}_{xy} = 1-s^{\mathbf{b}}_{xy}$ must be greater than or equal to zero. Since $D_{xy}$ is an average of $d^{\mathbf{b}}_{xy}$ values, it too is non-negative.

\subparagraph{Identity of indiscernibles}

We must show that $D_{xy} = 0$ if and only if $x=y$. 

First, suppose $x=y$. Then for all $\mathbf{b}$, $T^{\mathbf{b}}(x)=T^{\mathbf{b}}(y)$, and so  $|T^{\mathbf{b}}(x) \cap T^{\mathbf{b}}(y)| = |T^{\mathbf{b}}(x) \cup T^{\mathbf{b}}(y)|$. Then for all $\mathbf{b}$ (regardless of whether or not $\mathbf{b}$ ever contains $x$ or $y$), $s^{\mathbf{b}}_{xy}=1$ and $d^{\mathbf{b}}_{xy}=0$, and thus $D_{xy}$ is also zero.

To show the reverse implication, suppose that $x\ne y$. Then without loss of generality we can say that there is some time point $t$ for which $x(t) > y(t)$.

Suppose that there exists an observation $z\in X$ such that $z(t)<x(t)$, and consider the band $\mathbf{b}^*$ defined by $z$ and $y$. At $t$, $y$ is in $\mathbf{b}^*$; but $x(t)$ is strictly greater than both $y(t)$ and $z(t)$, so $x$ is not in $\mathbf{b}^*$ at $t$. Then  $|T^{\mathbf{b}^*}(x) \cap T^{\mathbf{b}^*}(y)| < |T^{\mathbf{b}^*}(x) \cup T^{\mathbf{b}^*}(y)|$, so $s^{\mathbf{b}^*}_{xy} < 1$ and $d^{\mathbf{b}^*}_{xy} > 0$. Since there is now a positive contribution to $D_{xy}$, we know that $D_{xy} > 0$.

Now suppose there exists an observation $z\in X$ such that $z(t)\ge x(t)$, and consider the band $\mathbf{b}^*$ defined by $z$ and $x$. At $t$, $x$ is in $\mathbf{b}^*$ but $y$ is not (since $y(t)$ is less than both $z(t)$ and $x(t)$). By similar reasoning to the above, we see that $D_{xy} > 0$. Thus $D_{xy}$ is 0 only if $x=y$.

\subparagraph{Symmetry}

We must show that $D_{xy} = D_{yx}$. This is evident from inspection of the definition;  $\frac{|T^{\mathbf{b}}(x) \cap T^{\mathbf{b}}(y)|} {|T^{\mathbf{b}}(x) \cup T^{\mathbf{b}}(y)|} = \frac{|T^{\mathbf{b}}(y) \cap T^{\mathbf{b}}(x)|} {|T^{\mathbf{b}}(y) \cup T^{\mathbf{b}}(x)|}$, and $B_{xy} = B_{yx}$.

\subparagraph{Triangle inequality}

We must show that $D_{xz} \le D_{xy} + D_{yz}$. Note that since $x$, $y$, and $z$ were selected arbitrarily, it suffices to demonstrate the inequality for this configuration of points.

Let $B_{xyz}$ be the set of all bands that contain $x$ or $y$ or $z$ at any time, and let $O=|B_{xyz}|$. Note that we need only be concerned with bands in $B_{xyz}$; if a band never contains either $x$ or $y$, for example, it does not affect $D_{xy}$, and so bands that are not in $B_{xyz}$ have no effect on either side of the inequality we are trying to show. 

We also need to denote certain subsets of $B_{xyz}$. Specifically, let
\begin{itemize}
\item $B_{xz}$ be the set of all bands that contain $x$ or $z$ at any time, and
\item $B_{-xz}$ be the set $B_{xyz}\backslash B_{xz}$, that is, the set of bands that contain $y$ at some time but do not contain $x$ or $z$ at any time.
\end{itemize}
The sets $B_{xy}$, $B_{yz}$, and so on are defined accordingly. We use $O_{xy}$ to represent the cardinality of sets of the type $B_{xy}$. For convenience, let $p$ be the cardinality of $B_{xz}$, and $q$ be the cardinality of $B_{-xz}$.

Note that $B_{xz}$ and $B_{-xz}$ partition the set $B_{xyz}$, so any relevant band will fall into exactly one of these subsets. Also note that $O=p+q$.

Recall the definitions. The bandwise similarity of $x$ and $y$, $s^b_{xy}$, is:
\[
s^{\mathbf{b}}_{xy} = \begin{cases}
\frac{|T^{\mathbf{b}}_x \cap T^{\mathbf{b}}_y|}{|T^{\mathbf{b}}_x \cup T^{\mathbf{b}}_y|} & \mbox{if } |T^{\mathbf{b}}_x\cup T^{\mathbf{b}}_y| \ne 0 \\
1  & \mbox{if } |T^{\mathbf{b}}_x\cup T^{\mathbf{b}}_y| = 0
\end{cases}
\]

The bandwise distance of $x$ and $y$, $d^{\mathbf{b}}_{xy}$, is simply $1-s^{\mathbf{b}}_{xy}$ (so if neither $x$ nor $y$ is ever in the band $\mathbf{b}$, then $d^{\mathbf{b}}_{xy}=0$). Note that, considered for a particular $\mathbf{b}$, $d^{\mathbf{b}}_{xy}$ is the Jaccard distance and is itself a distance metric.

Finally, the overall distance of $x$ and $y$, $D_{xy}$, is the average of the bandwise distance scores over all valid bands (that is, bands that contain $x$ or $y$ at some time):
\begin{eqnarray*}
D_{xy} & = & O_{xy}^{-1} \sum_{\mathbf{b}\in B_{xy}} d^{\mathbf{b}}_{xy} \\
 & = & O_{xy}^{-1} \sum_{\mathbf{b}\in B_{xyz}} d^{\mathbf{b}}_{xy}
\end{eqnarray*}
since $d^{\mathbf{b}}_{xy}=0$ for any $\mathbf{b} \notin B_{xy}$.

Initially, we consider the case where $q=0$: that is, any band that contains observation $y$ at any time must also contain $x$ or $z$ at some time. Because $d^{\mathbf{b}}_{xz} \le d^{\mathbf{b}}_{xy} + d^{\mathbf{b}}_{yz}$ for any given $\mathbf{b}$, we have
\[  \sum_{\mathbf{b} \in {B}} d^{\mathbf{b}}_{xz} \le  \sum_{\mathbf{b} \in {B_{xyz}}} d^{\mathbf{b}}_{xy} +  \sum_{\mathbf{b} \in {B_{xyz}}} d^{\mathbf{b}}_{yz}. \]
Now, both $O_{xy}$ and $O_{yz}$ are necessarily less than or equal to $O$, and so the quantities $\frac{O}{O_{xy}}$ and $\frac{O}{O_{yz}}$ are both greater than or equal to one. Then we can multiply terms on the right-hand side of the inequality by these factors and preserve the inequality:
\[  \sum_{\mathbf{b} \in {B_{xyz}}} d^{\mathbf{b}}_{xz} \le  \frac{O}{O_{xy}}\sum_{\mathbf{b} \in {B_{xyz}}} d^{\mathbf{b}}_{xy} + \frac{O}{O_{yz}} \sum_{\mathbf{b} \in {B_{xyz}}} d^{\mathbf{b}}_{yz}. \]
Then we have:
\[
\frac{1}{O} \sum_{\mathbf{b} \in {B_{xyz}}} d^{\mathbf{b}}_{xz}  \le  \frac{1}{O_{xy}}\sum_{\mathbf{b} \in {B_{xyz}}} d^{\mathbf{b}}_{xy} + \frac{1}{O_{yz}} \sum_{\mathbf{b} \in {B_{xyz}}} d^{\mathbf{b}}_{yz},
\]
but since $q=0$, we know that $O=O_{xz}$, so the inequality becomes
\[
\frac{1}{O_{xz}} \sum_{\mathbf{b} \in {B_{xyz}}} d^{\mathbf{b}}_{xz}  \le  \frac{1}{O_{xy}}\sum_{\mathbf{b} \in {B_{xyz}}} d^{\mathbf{b}}_{xy} + \frac{1}{O_{yz}} \sum_{\mathbf{b} \in {B_{xyz}}} d^{\mathbf{b}}_{yz},
\]
which by definition tells us that $D_{xz} \le D_{xy} + D_{yz}$, as desired.

Now, suppose that $q\ne 0$. We can list the bands in $B_{xyz}$ as follows:
\[B_{xyz}=
\begin{bmatrix}
B_{xz} \\
B_{-xz}
\end{bmatrix}
\]

We then consider an expanded set of (non-unique) bands, in which we include $p$ copies of $B_{xz}$ followed by $p$ copies of $B_{-xz}$:
\[
\tilde{B} :=
\begin{bmatrix}
B_{xz} \\
\vdots \\
B_{xz} \\
B_{-xz} \\
\vdots \\
B_{-xz}
\end{bmatrix}
\]

And finally a set of extended bands, formed by appending a band from $B_{xz}$ to each band in $\tilde{B}$:
\[
\hat{B} :=
\begin{bmatrix}
B_{xz} & B_{xz} \\
\vdots & \vdots \\
B_{xz} & B_{xz} \\
B_{-xz} & B_{xz} \\
\vdots & \vdots \\
 & B_{xz} \\
\vdots & \vdots \\
B_{-xz} & B_{xz}
\end{bmatrix}
\]
The first section of $\hat{B}$ consists of $p$ copies of $B_{xz}$, each reduplicated to form a band of twice the original length. In the second section, the first half of each band is some band drawn from $B_{-xz}$, and the second half of each band is drawn from $B_{xz}$. There are a total of $pq$ bands in this section; each band in $B_{-xz}$ appears (as the first half of a band) $p$ times, and each band in $B_{xz}$ appears (as the second half of a band) $q$ times.

Now we examine the behavior of the bandwise distance $d^{\mathbf{b}}_{xz}$ on these sets. First, note that $p\sum_{\mathbf{b}\in B_{xyz}} d^{\mathbf{b}}_{xz} = \sum_{ \tilde{\mathbf{b}} \in \tilde{B} } d^{ \tilde{\mathbf{b}} }_{xz}$ since each band in $B_{xyz}$ appears $p$ times in $\tilde{B}$.

Now, consider some band $\mathbf{b}^+ \in \hat{B}$. Suppose it is from the first section; that is, it has the form $[\begin{smallmatrix} \mathbf{b} & \mathbf{b} \end{smallmatrix}]$, where $\mathbf{b}$ is a band in $B_{xz}$. We have:
\begin{align*}
d^{\mathbf{b}^+}_{xz} & = 1 - \frac{|T^{\mathbf{b}^+}_x \cap T^{\mathbf{b}^+}_z|}{|T^{\mathbf{b}^+}_x \cup T^{\mathbf{b}^+}_z|} \\
 & = 1 - \frac{2|T^{\mathbf{b}}_x \cap T^{\mathbf{b}}_z|}{2|T^{\mathbf{b}}_x \cup T^{\mathbf{b}}_z|} && \text{as $\mathbf{b}^+$ contains two disjoint repetitions of $\mathbf{b}$} \\
 & = 1 - \frac{|T^{\mathbf{b}}_x \cap T^{\mathbf{b}}_z|}{|T^{\mathbf{b}}_x \cup T^{\mathbf{b}}_z|} \\
 & = d^{\mathbf{b}}_{xz}.
\end{align*}

Next suppose the band $\mathbf{b}^+$ has the form $\begin{bmatrix} \mathbf{b}^- & \mathbf{b} \end{bmatrix}$ where $\mathbf{b}$ is drawn from $B_{xz}$ and $\mathbf{b}^-$ is drawn from $B_{-xz}$. Then we have:
\begin{align*}
d^{\mathbf{b}^+}_{xz} & = 1 - \frac{|T^{\mathbf{b}^+}_x \cap T^{\mathbf{b}^+}_z|}{|T^{\mathbf{b}^+}_x \cup T^{\mathbf{b}^+}_z|} \\
 & = 1 - \frac{|T^{\mathbf{b}}_x \cap T^{\mathbf{b}}_z|}{|T^{\mathbf{b}}_x \cup T^{\mathbf{b}}_z|} = d^{\mathbf{b}}_{xz},
\end{align*}
as $\mathbf{b}^-$ contributes nothing to either $T^{\mathbf{b}^+}_x$ or $T^{\mathbf{b}^+}_z$, since neither $x$ nor $z$ ever appears in $\mathbf{b}^-$.

Each $\mathbf{b}\in B_{xz}$ appears $(p+q)$ times, extended in one or the other of these ways, in $\hat{B}$. In either case, $d^{\mathbf{b}^+}_{xz} = d^{\mathbf{b}}_{xz}$. So
\begin{eqnarray*}
\sum_{\mathbf{b}^+ \in \hat{B}} d^{\mathbf{b}^+}_{xz} & = & (p+q) \sum_{\mathbf{b}\in B_{xz}} d^{\mathbf{b}}_{xz} \\
 & = & (p+q) O_{xz} D_{xz}.
\end{eqnarray*}

Now we look at the observations $x$ and $y$, noting that the same reasoning will apply to the pair $y$ and $z$. As before, $p\sum_{\mathbf{b}\in B_{xyz}} d^{\mathbf{b}}_{xy} = \sum_{\tilde{\mathbf{b}}\in \tilde{B}} d^{\tilde{\mathbf{b}}}_{xy}$ since each band in $B_{xyz}$ appears $p$ times in $\tilde{B}$.

Consider a band $\mathbf{b}^+ \in \hat{B}$ of the form $\begin{bmatrix} \mathbf{b} & \mathbf{b} \end{bmatrix}$, where $\mathbf{b}\in B_{xz}$. We have:
\begin{align*}
d^{\mathbf{b}^+}_{xy} & = 1 - \frac{|T^{\mathbf{b}^+}_x \cap T^{\mathbf{b}^+}_y|}{|T^{\mathbf{b}^+}_x \cup T^{\mathbf{b}^+}_y|} \\
 & = 1 - \frac{2|T^{\mathbf{b}}_x \cap T^{\mathbf{b}}_y|}{2|T^{\mathbf{b}}_x \cup T^{\mathbf{b}}_y|} \\
 & = 1 - \frac{|T^{\mathbf{b}}_x \cap T^{\mathbf{b}}_y|}{|T^{\mathbf{b}}_x \cup T^{\mathbf{b}}_y|} \\
 & = d^{\mathbf{b}}_{xy} && \text{provided $|T^{\mathbf{b}}_x\cup T^{\mathbf{b}}_y|\ne 0$.}
\end{align*}
If $|T^{\mathbf{b}}_x\cup T^{\mathbf{b}}_y| = 0$, on the other hand, then neither $x$ nor $y$ is ever in $\mathbf{b}$, and therefore neither is ever in $\mathbf{b}^+$. In this case, then, $d^{\mathbf{b}^+}_{xy} = 0 = d^{\mathbf{b}}_{xy}$.

Now consider $\mathbf{b}^+\in \hat{B}$ of the form $\begin{bmatrix} \mathbf{b}^- & \mathbf{b} \end{bmatrix}$ where $\mathbf{b}\in B_{xz}$ and $\mathbf{b}^- \in B_{-xz}$. We have:
\begin{align*}
d^{\mathbf{b}^+}_{xy} & = 1 - \frac{|T^{\mathbf{b}^+}_x \cap T^{\mathbf{b}^+}_y|}{|T^{\mathbf{b}^+}_x \cup T^{\mathbf{b}^+}_y|} \\
 & = 1 - \frac{|T^{\mathbf{b}^-}_x \cap T^{\mathbf{b}^-}_y| + |T^{\mathbf{b}}_x \cap T^{\mathbf{b}}_y|}{|T^{\mathbf{b}^-}_x \cup T^{\mathbf{b}^-}_y| + |T^{\mathbf{b}}_x \cup T^{\mathbf{b}}_y|} && \text{as $\mathbf{b}^-$ and $\mathbf{b}$ are appended disjointly} \\
 & \le 1 && \text{since this fraction is $\ge 0$.}
\end{align*}

But note that $d^{\mathbf{b}^-}_{xy} = 1 - \frac{|T^{\mathbf{b}^-}_x \cap T^{\mathbf{b}^-}_y|}{|T^{\mathbf{b}^-}_x \cup T^{\mathbf{b}^-}_y|}$.  Because $\mathbf{b}^- \in B_{-xz}$, we know $T^{\mathbf{b}^-}_y \ne \emptyset$, so that the denominator of this fraction is not zero; but the numerator is zero since $T^{\mathbf{b}^-}_x = \emptyset$. Then the fraction is zero, and $d^{\mathbf{b}^-}_{xy} = 1$. Thus $d^{\mathbf{b}^+}_{xy} \le d^{\mathbf{b}^-}_{xy}$.

Every $\mathbf{b}^+ \in \hat{B}$ maps to a band in $\tilde{B}$ of one or the other of these types. Therefore, $\sum_{\mathbf{b}^+ \in \hat{B}} d^{\mathbf{b}^+}_{xy} \le \sum_{\tilde{\mathbf{b}} \in \tilde{B}} d^{\tilde{\mathbf{b}}}_{xy}$.

Combining these results, we have:
\begin{align*}
(p+q)O_{xz}D_{xz} & = (p+q)\sum_{\mathbf{b} \in B_{xyz}} d^{\mathbf{b}}_{xz} \\
 & = \sum_{\mathbf{b}^+ \in \hat{B}} d^{\mathbf{b}^+}_{xz} \\
 & \le \sum_{\mathbf{b}^+ \in \hat{B}} \big( d^{\mathbf{b}^+}_{xy} + d^{\mathbf{b}^+}_{yz} \big) && \text{by triangle ineq. on each $\mathbf{b}^+$} \\
 & =  \sum_{\mathbf{b}^+ \in \hat{B}} d^{\mathbf{b}^+}_{xy} +  \sum_{\mathbf{b}^+ \in \hat{B}} d^{\mathbf{b}^+}_{yz} \\
 & \le  \sum_{\tilde{\mathbf{b}} \in \tilde{B}} d^{\tilde{\mathbf{b}}}_{xy} + \sum_{\tilde{\mathbf{b}} \in \tilde{B}}  d^{\tilde{\mathbf{b}}}_{yz} \\
 & = p \big( \sum_{{\mathbf{b}} \in {B_{xyz}}} d^{{\mathbf{b}}}_{xy} + \sum_{{\mathbf{b}} \in {B_{xyz}}}  d^{{\mathbf{b}}}_{yz} \big) \\
 & = p (O_{xy}D_{xy} + O_{yz}D_{yz}) \\
 & \le p \max(O_{xy},O_{yz}) (D_{xy} + D_{yz}),
\end{align*}
and so
\[ D_{xz} \le \frac{p \max(O_{xy}, O_{yz})}{(p+q)O_{xz}} (D_{xy} + D_{yz}).\]

But
\begin{align*}
\frac{p \max(O_{xy}, O_{yz})}{(p+q)O_{xz}} & = \frac{\max(O_{xy},O_{yz})}{p+q} && \text{as $p=O_{xz}$} \\
 & = \frac{max(O_{xy},O_{yz})}{O} \\
 & \le 1,
\end{align*}
since $B_{xy}$ and $B_{yz}$ are both subsets of $B_{xyz}$. Then the above inequality becomes
\[ D_{xz} \le 1 (D_{xy} + D_{yz})\]
and so
\[ D_{xz} \le D_{xy} + D_{yz}.\]

%%%%%%%%%%%%%%%%
%%%% BIBLIOGRAPHY %%%%

\bibliography{tupper_band_distance}

\end{document}